\documentclass[11pt,twoside]{article}

\usepackage{asp2004}
\usepackage{epsf}
\usepackage{epsfig}
\usepackage{lscape}
\usepackage{natbib}

\begin{document}
\title{The subdwarf B + white dwarf binary KPD~1930+2752\\
       a Supernova Type Ia progenitor candidate}   
\author{S. Geier$^1$, S. Nesslinger$^1$, U. Heber$^1$, N. Przybilla$^1$, R. Napiwotzki$^2$ and R.-P. Kudritzki$^3$}   
\affil{$^1$ Dr.--Remeis--Sternwarte, Institute for Astronomy, University Erlangen-Nuremberg, Sternwartstr. 7, 96049 Bamberg, Germany\\
$^2$ Centre of Astrophysics Research, University of Hertfordshire, College Lane, Hatfield AL10 9AB, UK\\
$^3$ Institute for Astronomy, University of Hawaii, 2680 Woodlawn Drive, Honolulu, HI 96822, USA}


\begin{abstract} 
The nature of the progenitors of type Ia supernovae is still under controversial
debate. KPD~1930+2752 is one of the best SN Ia progenitor candidates known
today. The object is a double degenerate system consisting of a
subluminous B star and a massive white dwarf. \citet{maxted}
conclude that the system mass exceeds the Chandrasekhar mass. This
conclusion, however, rests on the assumption that the sdB mass is
$0.5\,M_{\odot}$. However, recent binary population synthesis calculations
suggest that the mass of an sdB star may range from $0.3\,M_{\odot}$ to more
than $0.7\,M_{\odot}$. 
It is therefore important to measure the mass of the sdB star simultaneously with that
of the white dwarf.
Since the rotation of the sdB star is tidally locked to the orbit the 
inclination of the system can 
be constrained. An analysis of the ellipsoidal variations in the 
light curve allows to tighten the constraints derived from spectroscopy.
We derive the mass-radius relation for the sdB star from 
a quantitative spectral analysis.
The projected rotational 
velocity is determined for the first time 
from high-resolution spectra. In addition a reanalysis of the published light curve 
is performed.
The atmospheric and orbital parameters are measured with unprecedented accuracy. In particular the projected rotational velocity \(v_{\rm rot}\sin{i} = 92.3 \pm 1.5~{\rm km\,s^{-1}}\) is determined.
The mass of the sdB is limited between \(0.45~M_{\odot}\) and \(0.52~M_{\odot}\). The total mass of the system ranges from
\(1.36~M_{\odot}\) to \(1.48~M_{\odot}\) and hence is likely to exceed the Chandrasekhar mass. So KPD~1930+2752 qualifies as an excellent double degenerate supernova Ia progenitor candidate.
\end{abstract}


\section{Introduction}   

Supernovae of type Ia (SN~Ia) play an important role in the study of cosmic 
evolution. They are regarded as the best standard candles for the determination of the cosmological parameters. 
The nature of their progenitors is still under debate. 
There is general consensus that only the thermonuclear explosion of a 
white dwarf (WD) is compatible with the observed features of SN~Ia. 
A white dwarf has to accrete mass from a close companion to reach 
the Chandrasekhar limit of $1.4 \,M_{\rm \odot}$. Two main scenarios of 
mass transfer are currently under discussion. 
In the so-called single degenerate (SD) scenario \citep{whelan},
 the mass donating component is a red giant/subgiant, which fills its Roche 
 lobe and is continually transfering mass to the white dwarf. According to 
 the so-called double degenerate (DD) scenario \citep{iben} 
 the mass donating companion is a white dwarf, which eventually merges 
 with the primary due to orbital shrinkage caused by gravitational wave 
 radiation.
KPD~1930+2752 was identified as a pulsating subdwarf B star (sdBV). 
Multiperiodic variations with short periods and low amplitudes were detected. In addition to these
pulsations a strong variation at a much longer period of about \(4100 \, {\rm s}\) was found. 
This variation could be identified as an ellipsoidal deformation of the sdB 
star most likely caused by a massive companion. \citet{billeres} 
predicted the period of the binary to be two times the period of this 
brightness variation. This was proven by \citet{maxted}, 
who measured a radial 
velocity curve of KPD~1930+2752 which matched the proper period. The radial 
velocity amplitude combined with an assumption of the canonical mass for sdB 
stars \(M_{ \rm sdB}=0.5 \, M_{\odot}\) led to a lower limit for the mass of 
the system derived from the mass function. 
This lower limit \(M \geq 1.47\, M_{\odot} \) exceeded the Chandrasekhar 
mass of \(1.4 \, M_{\odot}\). The system is considered to become double degenerate when the subdwarf 
eventually evolves to a white dwarf. Orbital shrinkage caused by 
gravitational wave radiation will lead to a merger of the binary. 
Maxted et al. concluded that KPD~1930+2752 could be a good candidate for 
the progenitor of a Type Ia supernova.\\
This conclusion rests on the assumption that the mass of the primary is $0.5 \, M_{\odot}$. sdB stars have been identified as core helium burning stars at the blue end of the Horizontal Branch (Extreme Horizontal Branch, EHB). According to canonical stellar evolution the mass is fixed by the onset of the core helium flash to about half a solar mass \citep{heber}. However, binary population synthesis calculations \citep{han} suggest, that the mass range for sdB stars is much wider, ranging from $0.3\,M_{\odot}$ to more than $0.7\,M_{\odot}$. Therefore we aim at 
measuring the mass of the sdB star simultaneously with that
of the white dwarf.
Since the rotation of the sdB star is tidally locked to the orbit the 
inclination of the system can 
be constrained if the sdB radius and the projected rotational velocity 
can be measured with high precision. An analysis of the ellipsoidal variations in the 
light curve allows to tighten the constraints derived from spectroscopy.
We derive the mass-radius relation for the sdB star from 
a quantitative spectral analysis.
The projected rotational 
velocity is determined for the first time 
from high-resolution spectra. In addition a reanalysis of the published light curve 
is performed.

\section{Observations and Data Analysis}

With the \(10 \, {\rm m}\) Keck I Telescope at the Mauna Kea Observatory two 
hundred high-resolution spectra were obtained using the High Resolution Echelle Spectrometer (HIRES). 
Two spectra were taken with the ESO Very Large Telescope UT2 (Kueyen) equipped 
with the UV-Visual Echelle Spectrograph (UVES). 
Additional 150 low-resolution spectra were obtained with the \(2.2 \, {\rm m}\) Telescope at 
the Calar Alto Observatory using the Calar Alto Faint Object 
Spectrograph (CAFOS).\\
The data was reduced using the ESO-MIDAS package.
The radial velocities were obtained by cross correlation with a model spectrum 
at rest wavelength. Combining with data from literature we derive: 
\(K=341 \pm 1 \,{\rm km\,s^{-1}}\), 
\(P=0.0950933 \pm 0.0000015~ {\rm d}\).\\
For the analysis of the spectra of KPD~1930+2752 we used LTE model 
grids for 10 times solar metallicity.
As the surface gravity is of utmost importance 
for our analysis we also calculated new grids of models and synthetic  
spectra to account for NLTE effects and metal line blanketing 
simultaneously. 
The mean parameter values from the model atmosphere fits are: $T_{\rm eff} = 35\,200 \pm 500\, {\rm K}$ and 
$\log g = 5.61 \pm 0.06\, {\rm dex}$.\\ 
The primary aim of the high-resolution time-series spectroscopy was to 
measure 
the projected rotational velocity of KPD~1930+2752 as accurately as possible.
For this purpose the spectra obtained with HIRES and UVES were used.
The projected rotational velocity was measured by convolving a synthetic spectrum calculated 
from the best fit model atmosphere with a rotational broadening ellipse for appropriate 
$v_{\rm rot}\sin i$ (see figure above). The result is very accurate: 
\(v_{\rm rot}\sin{i} = 92.3 \pm 1.5 \, {\rm km\,s^{-1}}\)

\section{Mass and Inclination}

KPD~1930+2752 is obviously affected by the gravitional forces of 
the companion, demonstrated by its ellipsoidal deformation.
Since the period of the photometric variations is exactly half the period of 
the radial velocity variations the 
rotation of the sdB star is tidally locked to the orbit.
Having determined the gravity and projected rotational velocity, we have 
three equations at hand that constrain the system, with the sdB mass 
\(M_{\rm sdB}\) being the only free parameter.
Besides the mass function
\begin{equation} 
	f(M_{\rm sdB},M_{\rm comp})=\frac{M_{\rm comp}^{3}(\sin{i})^{3}}{(M_{\rm sdB}+M_{\rm comp})^{2}}=\frac{K_{\rm sdB}^{2}P}{2\pi G}
	\label{massfunk}
\end{equation}
these are
\begin{displaymath} 
	\qquad \qquad \qquad \qquad  \sin{i}=\frac{v_{\rm rotsini}P}{2\pi R} \qquad {\rm and} \qquad R=\sqrt{\frac{M_{\rm sdB}G}{g}} \qquad \qquad 
        \qquad  {\rm (2,3)}
\end{displaymath}
With \(\log \, g\) obtained from the model atmosphere analysis, the radius of 
the star $R$ was calculated using the standard mass-radius relation 
(Eq. 3). 
Together with the orbital period of the system \(P\) and the projected 
rotational velocity $v_{\rm rotsini} = v_{\rm rot}\sin{i}$ the inclination of 
the system $\sin{i}$ was derived for different values of the sdB mass. 
Because the rotation is tidally locked to the orbit, the rotational period of 
the sdB equals the orbital period of the system. Therefore the absolute value 
of the rotational velocity could be calculated. From 
the measured projected rotational velocity the inclination of the system 
follows (Eq. 2). With the sdB mass as free parameter, the 
measured radial velocity semiamplitude $K_{\rm sdB}$ and orbital 
period $P$ the mass function was solved 
numerically (Eq. \ref{massfunk}) to derive the mass of the companion 
$M_{\rm comp}$ and calculate the total mass of the binary. The fact that 
\(\sin{i}\) cannot exceed unity gave a lower limit for 
the mass of the sdB $M_{\rm sdB}\geq0.45\,M_{\odot}$. 
The total mass of the system exceeds the 
Chandrasekhar limit for almost all assumptions of \(M_{\rm sdB}\). 
The inclination angle of the system is close to \(90^{\circ} \) implying that 
KPD~1930+2752 may be an eclipsing binary.

\begin{figure}[h!]

\label{fig}
\plottwo{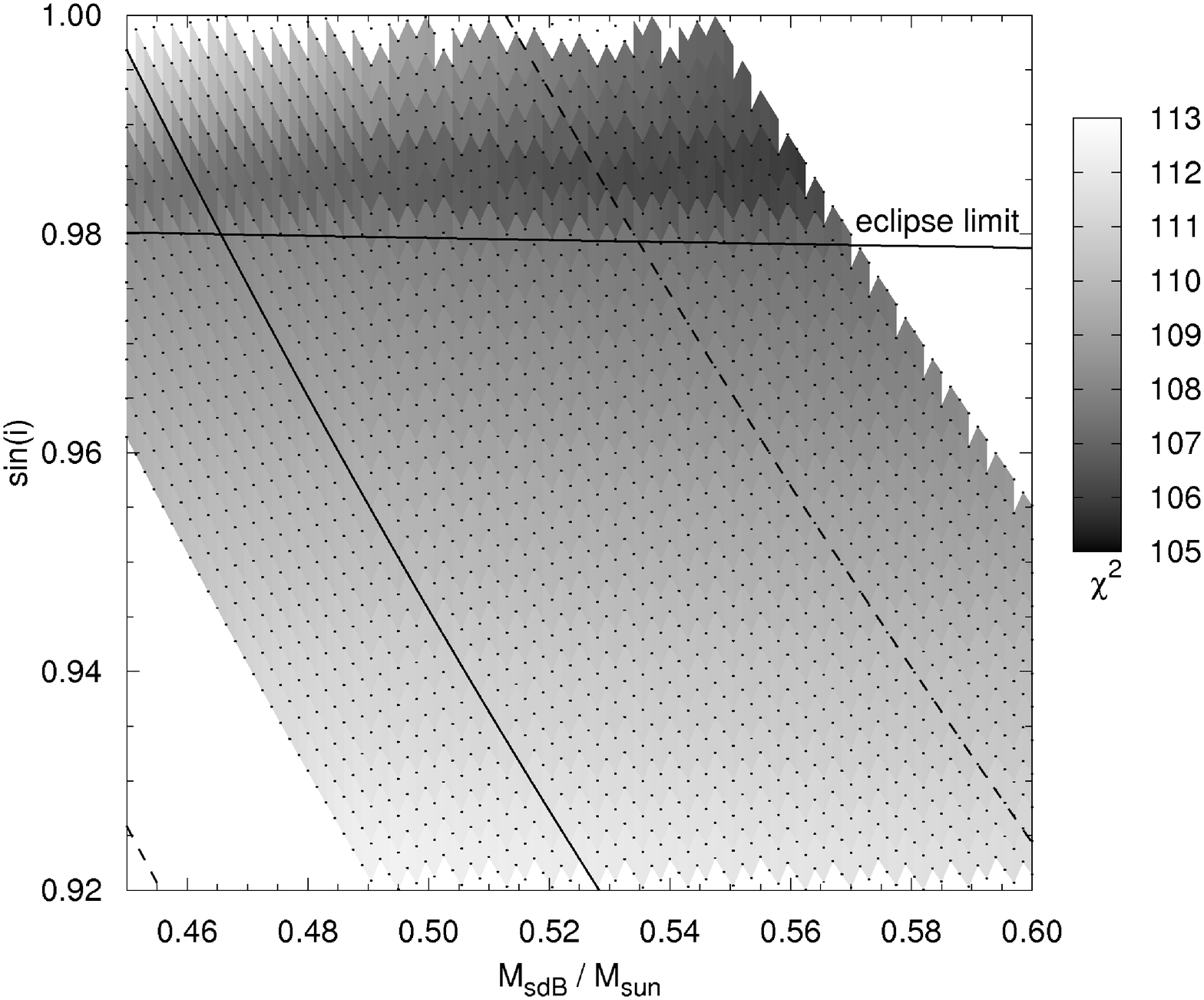}{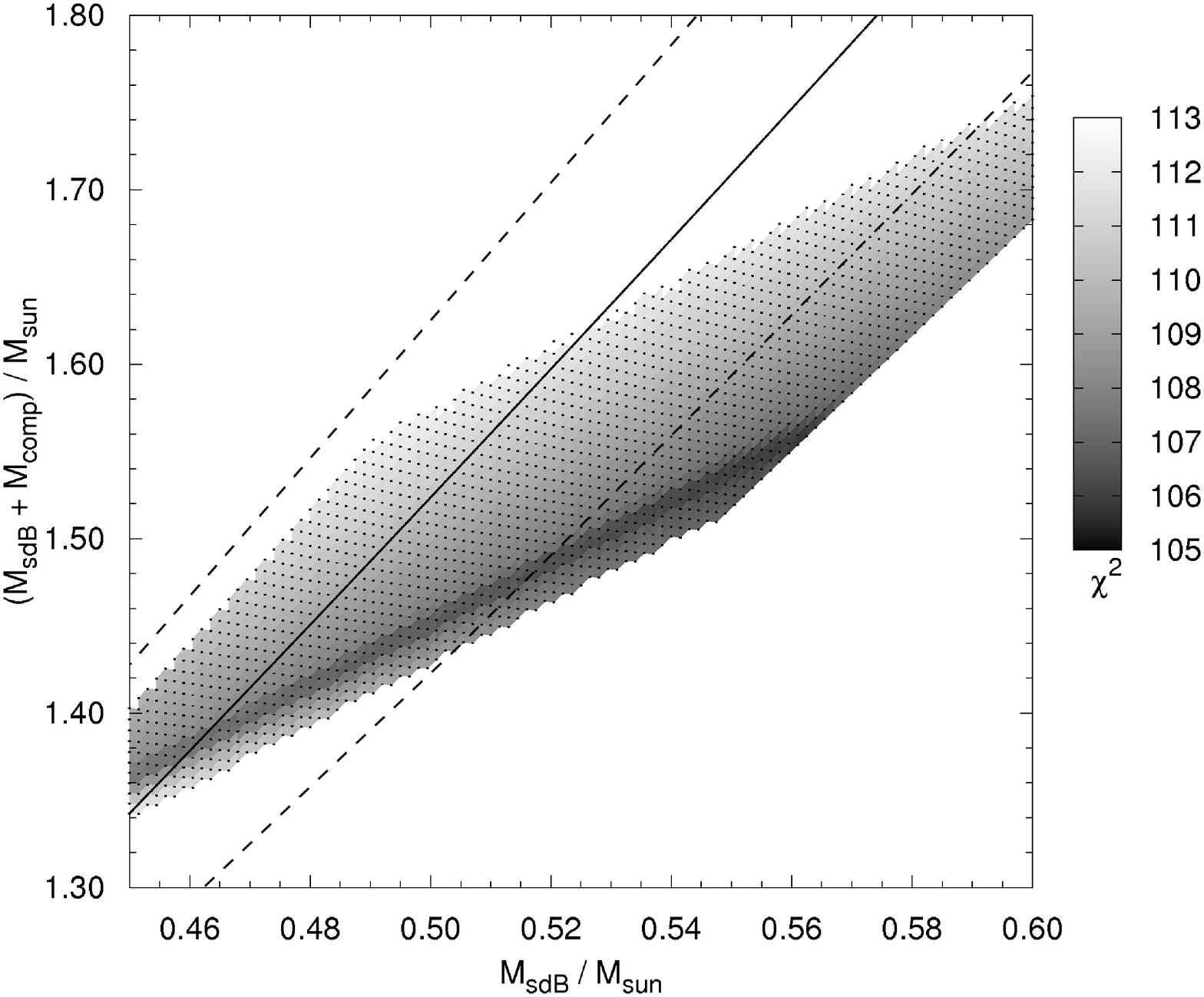}
\caption{The left figure shows the $\sin{i}$ as a function of sdB mass. The full drawn line
marks the solution derived from spectroscopy, the dashed lines the associated errors in $\sin{i}$.
Each dot denotes the location of a parameter set for which a light curve has been calculated. The background shading corresponds to the quality of the fit -- darker shading implying better $\chi^2$. The similar figure in the
right shows the secondary mass as a function of the sdB mass.}

\end{figure}

\section{Constraints by Photometry}

The light curve obtained by \citet{billeres} shows 
ellipsoidal variations. We used this information to further constrain the 
parameters of the KPD~1930+2752 system. We employed the light curve synthesis and
solution code MORO which is based on the model by \citet{wilson}. A very fine grid of synthetic light curves was constructed,
containing more than 76000 parameter combinations. Indeed, a significant overlap 
between the spectroscopically
determined parameter range and a set of very good light curve fits could be
found. Most notable is the coincidence of the best fits with
the region of eclipse ($i\approx 80^{\circ}$) (see Fig. \ref{fig}). 
If we combine the results of the photometric analysis with the mass-inclination 
relation derived from spectroscopy, the sdB mass is 
constrained to a very narrow range of
$M_{\rm sdB} = 0.45 - 0.52\,M_{\odot}$, corresponding to a 
total mass of $1.36 - 1.48\,M_{\odot}$ (cf. Fig. \ref{fig}).





\end{document}